\documentclass[%
 reprint,
 amsmath,amssymb,
 aps,
]{revtex4-1}

\usepackage{graphicx}
\usepackage{dcolumn}
\usepackage{bm}

\begin{document}

\preprint{APS/123-QED}

\title{Distribution of the magnetic environment in charge ordered La$_{1.885}$Sr$_{0.115}$CuO$_4$}


\author{A.\ Arsenault$^{1}$} 
\author{T.\ Imai$^{1}$} 
\author{P.\ M.\ Singer$^{2}$}
\author{K.\ M.\ Suzuki$^{3}$}
\author{M.\ Fujita$^{3}$} 
\affiliation{$^{1}$Department of Physics and Astronomy, McMaster University, Hamilton, Ontario L8S4M1, Canada}
\affiliation{$^{2}$Department of Chemical and Biomolecular Engineering, Rice University, 6100 Main St., Houston, TX 77005-1892}
\affiliation{$^{3}$Institute for Materials Research, Tohoku University, Sendai 980-8577, Japan}

\date{\today}

\begin{abstract}
We report the inverse Laplace transform (ILT) analysis of the $^{139}$La nuclear spin-lattice relaxation rate $1/T_1$ in charge ordered La$_{1.885}$Sr$_{0.115}$CuO$_4$ ($T_{charge} \sim 80$~K, $T_{c} \simeq T_{spin}^{neutron}=30$~K), and shed new light on its magnetic inhomogeneity.  We {\it deduce} the probability density function $P(1/T_{1})$ of the distributed $1/T_1$ ({\it i.e.} the histogram of distributed $1/T_1$) by taking the inverse Laplace transform of the experimentally observed nuclear magnetization recovery curve $M(t)$.  We demonstrate that spin freezing sets in {\it in some domains} precisely below the onset of charge order at $T_{charge}$, but their volume fraction grows only gradually toward $T_{c}$.  Nearly a half of the sample volume exhibits properties expected for canonical high $T_c$ cuprates without charge order even near $T_c$.  Our findings explain why charge order does not suppress $T_c$ of La$_{1.885}$Sr$_{0.115}$CuO$_4$ as significantly as in La$_{1.875}$Ba$_{0.125}$CuO$_4$.
\end{abstract}

\maketitle


\section{\label{sec:level1} Introduction}
Recent advances in X-ray diffraction techniques led to successful detection of charge order Bragg peaks in La$_{2-x}$Sr$_{x}$CuO$_4$ ($x \sim 1/8$) below $T_{charge} \sim 80$~K \cite{Croft,Thampy, WenNatComm2019}.  The confirmation finally settled the old controversy stemming from our earlier reports that the unusual NMR anomalies identified at the charge order transition of La$_{1.48}$Nd$_{0.4}$Sr$_{0.12}$CuO$_4$ are shared with  La$_{1.88}$Sr$_{0.12}$CuO$_4$, La$_{1.875}$Ba$_{0.125}$CuO$_4$, and La$_{1.68}$Eu$_{0.2}$Sr$_{0.12}$CuO$_4$, and hence all of these La214 type cuprates undergo a charge order transition at a comparable temperature \cite{HuntPRL1999, SingerPRB1999, HuntPRB2001,below12}.  In fact, the charge order phenomenon turns out to be more ubiquitous  across different classes of high $T_c$ cuprates \cite{Tranquada, Fujita, Fink, WuNature2011, AckerPRL2012, BlackburnPRL2013, BlancoPRL2013, AckerPRL2014, HuckerPRB2014, BlancoPRB2014, daSilvaNeto}, contrary to the belief held by many researchers two decades ago that charge order was merely an odd byproduct of the structural transition into the low temperature tetragonal (LTT) phase of La$_{1.48}$Nd$_{0.4}$Sr$_{0.12}$CuO$_4$.

The fundamental difficulty encountered in the experimental investigation of charge order in the La214 family is that the amplitude of the charge density modulation is extremely small.  For the NMR investigation, there is an additional challenge: charge order triggers glassy spin freezing within charge ordered domains of the La214 family \cite{TranquadaPRB59}, and begins to locally enhance the NMR relaxation rates with a wide distribution \cite{HuntPRL1999, SingerPRB1999, HuntPRB2001}.  Since $^{63}$Cu NMR relaxation rates in high $T_c$ cuprates are generally enhanced by the Cu-Cu super-exchange interaction even without charge order, mild enhancement of the low frequency Cu spin fluctuations (and hence the NMR relaxation rates) by a factor of $\sim 3$ makes the $^{63}$Cu NMR signal detection difficult, i.e. signal intensity wipe out \cite{HuntPRL1999, SingerPRB1999, HuntPRB2001, ImaiPRB2017, ImaiPRB2019}.  In fact, our recent single crystal $^{63}$Cu NMR measurements on La$_{1.885}$Sr$_{0.115}$CuO$_4$ \cite{ImaiPRB2017} confirmed that normally behaving $^{63}$Cu signals are gradually wiped out below $T_{charge}$.  The lost spectral weight is gradually transferred to a broad, wing-like signals with extremely fast NMR relaxation rates.  The latter originates from charge ordered domains with enhanced spin correlations, and can be detected down to $\sim 30$~K only with extremely fast separation time $\tau \sim 2$~$\mu$s between the 90 and 180 degree radio frequency pulses.  We note that such an experimental condition was not technically feasible in the 1990's.  

In contrast, $^{139}$La NMR signals are observable in the entire temperature range, because hyperfine couplings between $^{139}$La nuclear spins and Cu electron spins are weaker, and hence the $^{139}$La NMR relaxation rates are much slower.  The downside of the $^{139}$La NMR study is that the spin-lattice relaxation rate $1/T_1$ in the charge ordered state has a large distribution due to the glassy nature of spin freezing below $T_{charge}$.  The NMR community did not have an effective tool to deal with such a distribution to make concrete statements about $1/T_1$ results.  In fact, two different types of $^{63}$Cu NMR signals are observed for La$_{1.885}$Sr$_{0.115}$CuO$_4$ with different $1/T_1$ \cite{ImaiPRB2017}, indicating the presence of two different types of domains, but the corresponding $^{139}$La NMR signals are superposed \cite{ArsenaultPRB2018}.  Then how shall we discern their $1/T_1$?  We recently tried to circumvent the difficulty by assuming that two distinct components exist in the $^{139}$La nuclear spin-lattice relaxation recovery curve $M(t)$ \cite{ArsenaultPRB2018}.  The relative fractions of the fast and slow $1/T_1$ contributions in $M(t)$ agree reasonably well with the relative intensities of the two different $^{63}$Cu NMR signals, but the double component fit is based on an assumption.

In this paper, we re-examine the $1/T_1$ process at the $^{139}$La sites based on the inverse Laplace transform (ILT) analysis of the experimentally observed $M(t)$ \cite{SingerJCP2018, SingerPRB2019, TakahashiPRX2019}.  The ILT allows us to {\it deduce} the probability density distribution $P(1/T_{1})$ for distributed $1/T_1$ (i.e. the histogram of $1/T_{1}$ distribution) without making any assumptions on the functional form of $M(t)$.  If $1/T_{1}$ is distributed around one value, $P(1/T_{1})$ will have one peak.  But $P(1/T_{1})$ will exhibit a double peak structure if $1/T_{1}$ is distributed about two distinct values, as we demonstrated elsewhere based on a simple example (see Fig.\ 1 of \cite{SingerPRB2019}).  In what follows, we show that $1/T_{1}$ in charge ordered La$_{1.885}$Sr$_{0.115}$CuO$_4$ indeed develops asymmetric distribution, indicative of the presence of two different types of domains below $T_{charge}$.  We establish that charge order does not uniformly affect the entire CuO$_2$ planes, and the volume fraction of the charge ordered domains grows only progressively toward $T_{c}=30$~K.

\section{\label{sec:level1} Experimental }
We grew the La$_{1.885}$Sr$_{0.115}$CuO$_4$ single crystal  at Tohoku using the traveling solvent floating zone method \cite{Kimura}.  We annealed the crystal in flowing oxygen gas atmosphere at 900~$^{\circ}$C for 72 hours.  The crystal used for the present study is the same piece as in our previous studies \cite{ImaiPRB2017, ArsenaultPRB2018}, and has the approximate dimensions of 3~mm $\times$ 3~mm $\times$ 1~mm.  We determined the superconducting critical temperature $T_{c} = 30$~K based on the magnetization measured with a superconducting quantum interference device (SQUID).  La$_{1.885}$Sr$_{0.115}$CuO$_4$ is known to enter the spin ordered phase below $T_{spin}^{neutron} \simeq T_{c}$ at the fast measurement time scale of elastic neutron scattering \cite{Kimura}, and the residual spin fluctuations become static at the slower measurement time scale of $\mu$SR below $T_{spin}^{\mu SR} \sim 15$~K \cite{Kumagai, Savici}.

We conducted all the $^{139}$La NMR measurements at McMaster based on standard pulsed NMR techniques in an external magnetic field of 9~T.  See \cite{ArsenaultPRB2018} for the NMR lineshapes, representative $M(t)$ measured for the $I_{z} = +1/2$ to -1/2 central transition based on an inversion recovery technique, and the single and double component stretched fits.   We have repeated some of the $M(t)$ measurements between 30~K and $T_{charge}$ using longer spin echo recycling time up to $t_{recycle}\sim 5$~s to ensure that we do not overlook the longest components of the distributed $1/T_1$ in $P(1/T_{1})$.  We note that the relaxation rate $1/T_{1}^{str}$ determined from the stretched exponential fit hardly changes even if one uses $t_{recycle}$ that is somewhat too short, but $P(1/T_{1})$ would lose longer components.  In addition,  the resolution of the ILT curve $P(1/T_{1})$ depends on the signal to noise ratio of $M(t)$, as discussed in the Supplementary Materials of  ref. \cite{SingerPRB2019}.  Accordingly, the noise level in the $M(t)$ curve should be kept as small as possible (ideally, $\sim 0.1$\%). 

\section{\label{sec:level1} Results and discussions}
The inverse Laplace transform (ILT) consists of fitting the measured recovery curve $M(t)$ to a sum of exponentials with decay rate $1/T_{1,j}$ and the probability density $P(1/T_{1,j}) \geq 0$ \cite{JohnstonPRL2005, SingerJCP2018, SingerPRB2019}.  More specifically, the ILT consists of inverting the following equation for $P(1/T_{1,j})$:
\begin{equation}
M(t) = \sum_{i,j}\left[1 - 2 p_i\exp(-q_i t/T_{1,j})\right] P(1/T_{1,j}),
\label{eq:ILT}
\end{equation}
where the summation $\sum_{j} P(1/T_{1j}) = M_0$ is the saturated value of the magnetization at very long delay times $t \gg T_1$.  For mathematical clarity, we assumed that inversion is perfect.  The coefficients $\{ p_{i} \}= \{1/84,3/44,75/364,1225/1716\}$ (where $\sum_{i=1}^4 p_i = 1$) and $\{q_{i}\} = \{1,6,15,28\}$ are theoretically calculated and fixed for each normal mode in the recovery of nuclear spin $I=7/2$ \cite{Andrew, Narath}.  Given the large distribution in $P(1/T_{1,j})$ at low temperatures in the present case, the width of the $1/T_{1,j}$ bins in the ILT are chosen to be equally spaced on a logarithmic scale, $\Delta_{P} = Log_{10}(1/T_{1,j+1}) - Log_{10}(1/T_{1,j}$) = 0.03213.  $\sum_j P(1/T_{1,j}) \Delta_{P} = 1$ is normalized to one so that $P(1/T_{1,j})$ represents the probability density for the relaxation rate to have the particular value $1/T_{1,j}$.  Inverting for $P(1/T_{1,j})$ in Eq. (\ref{eq:ILT}) is an ill-conditioned problem, and therefore requires Tikhonov regularization (i.e. a smoothing factor) for a solution \cite{Venkataramanan2002,SongJMR2002}.  We refer readers to Supplemental Materials of \cite{SingerPRB2019} for additional mathematical background and the optimization procedures of the ILT.   

\begin{figure}
	\centering
	\includegraphics[width=3.2in]{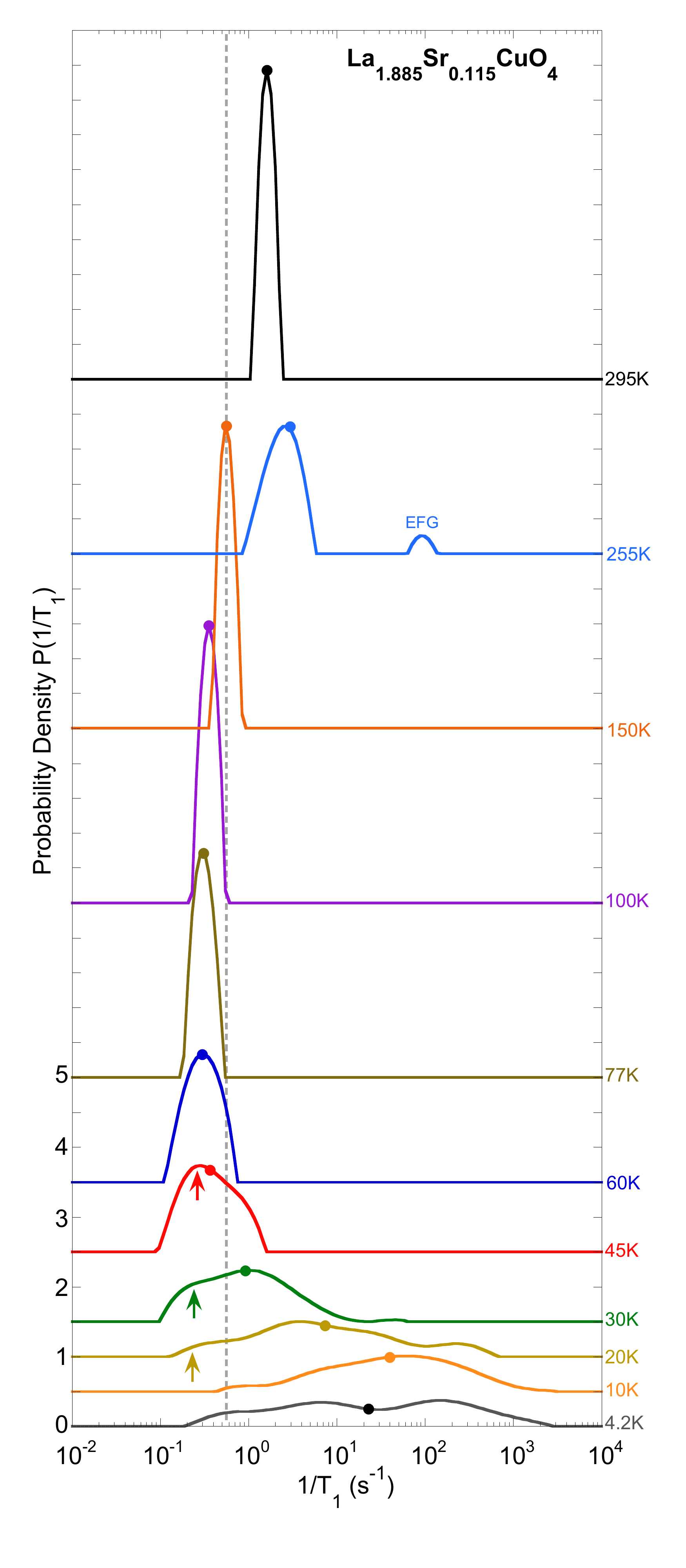}
	\caption[ILT]{Representative results of the probability density $P(1/T_{1})$ deduced with  ILT by numerically inverting experimentally observed $M(t)$.  For clarity, the origin is shifted vertically above 4.2~K.  Filled bullets mark $P(1/T_{1}^{lm})$ at the log-mean value of $1/T_{1}^{lm}$, the center of gravity of the $P(1/T_{1})$ curve on a log scale.  A small split-off peak marked as EFG at 255~K arises from the slow fluctuations of the EFG at the high temperature tetragonal to low temperature orthorhombic structural phase transition \cite{SingerPRB2019}.  Upward arrows show the peak (45~K) and shoulder (30~K and 20~K) overlooked by the conventional analysis based on stretched exponential fit.  Dashed vertical line is the cut off at $1/T_{1}=0.55$~s$^{-1}$ for the canonical superconducting domains, see the main text for details.}
	\label{fig:lineshape}
\end{figure}

In Fig.\ 1, we summarize the $P(1/T_1)$ curves at representative temperatures.  The filled bullet superposed on each curve marks $P(1/T_{1}^{lm})$ at the log-mean value of the relaxation rate, $1/T_{1}^{lm}$.  $1/T_{1}^{lm}$ is the center of gravity of $1/T_{1}$ distribution on a logarithmic scale, and represents the spatially averaged value of $1/T_{1}$ in the entire sample.  For example, the red $P(1/T_1)$ curve deduced at 45~K indicates that $1/T_1$ is distributed between $\sim 0.09$~s$^{-1}$ and  $\sim 1.6$~s$^{-1}$; the most likely value is $1/T_{1} \simeq 0.29$~s$^{-1}$ as determined by the location of the peak marked by the red upward arrow; the average value $1/T_{1}^{lm} \simeq 0.37$~s$^{-1}$ (red filled bullet) is not at the peak, because $P(1/T_1)$ is asymmetrical.  

We also summarize $P(1/T_1)$ as a color contour map in Fig.\ 2.  The additional relaxation processes caused by the slow fluctuations of the electric field gradient (EFG) enhance $1/T_{1}$ near the structural phase transition from the high temperature tetragonal (HTT) to the low temperature orthorhombic (LTO) structure at $T_{HTT-LTO} \simeq 255$~K, accompanied by the broader distribution and a split off peak at $1/T_{1} \sim 10^{2}$~s$^{-1}$.  We found analogous results near the structural phase transitions of  La$_{1.875}$Ba$_{0.125}$CuO$_4$ \cite{SingerPRB2019}.  Otherwise, the distribution of $P(1/T_1)$ is narrow from 295~K down to $T_{charge}$, and the log-mean value $1/T_{1}^{lm}$ decreases smoothly.

\begin{figure}
	\centering
	\includegraphics[width=3.2in]{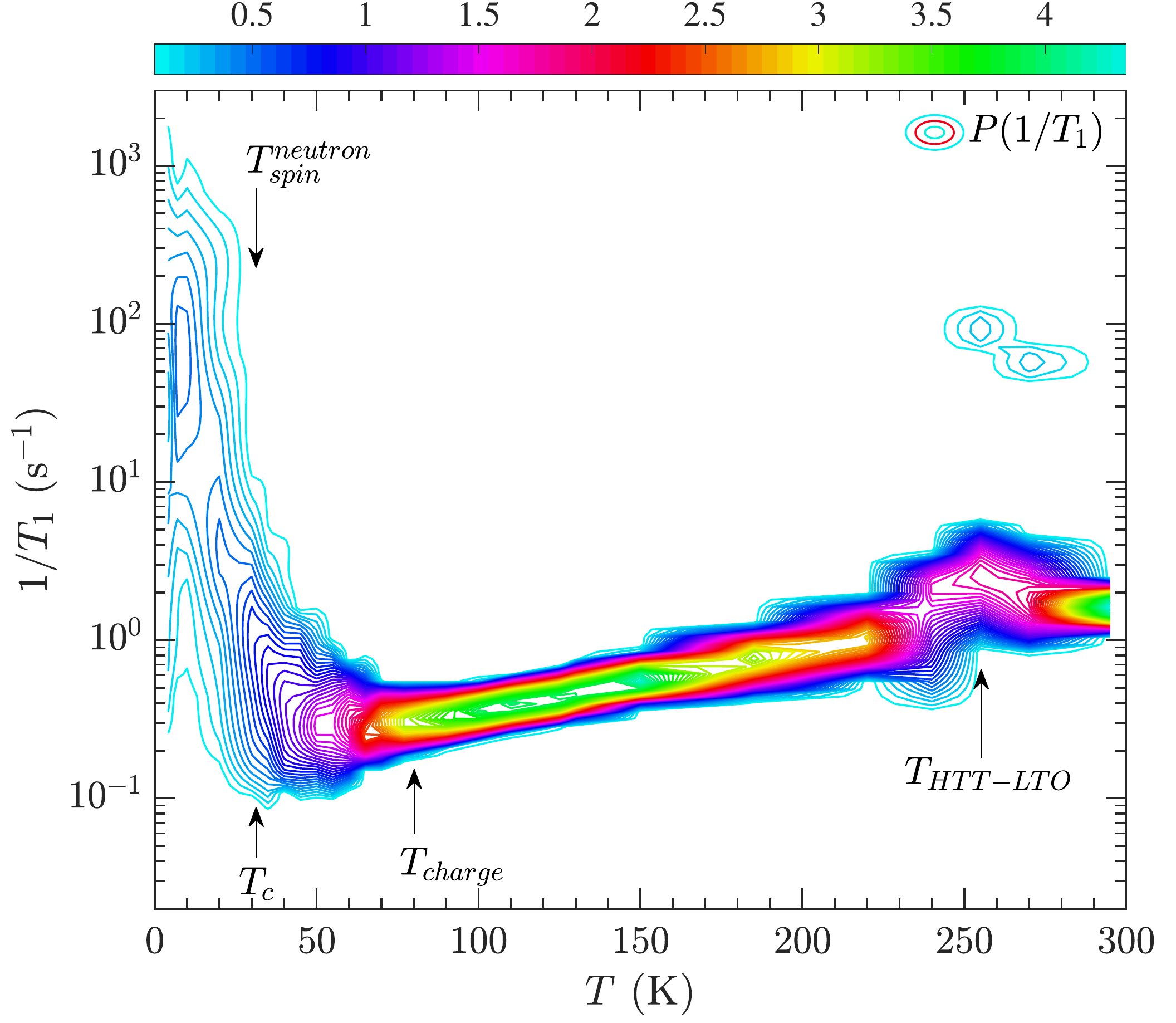}
	\caption{Contour map of $P(1/T_{1})$ (probability density) generated from ILT. Color bar scale is shown at the top of the figure.}
	\label{fig:f0}
\end{figure}

In Fig.\ 3(a), we summarize the temperature dependence of $1/T_{1}^{lm}$ (black filled bullets) and compare with $1/T_{1}^{str}$ estimated from the conventional single component stretched fit of $M(t)$ (green diamonds) \cite{ArsenaultPRB2018}.   $1/T_{1}^{str}$ agrees with the log-mean value $1/T_{1}^{lm}$, and hence the former is a good approximation of the latter.  Also summarized in Fig.\ 3(b) is the stretched exponent $\beta$ obtained from the stretched fit of $M(t)$ \cite{ArsenaultPRB2018}.  $\beta$ shows anti-correlation with the standard deviation $\sigma$ calculated for the distribution, $P(1/T_1)$.  That is, when $\beta$ decreases from 1 due to the distribution of $1/T_1$, $\sigma$ grows.  This makes sense, because deviation from $\beta = 1$ signals the extent of distribution in $1/T_1$, whereas $\sigma$ is the direct measure of the distribution of $1/T_1$.  These findings in Fig.\ 3 establish that the ILT results of $1/T_{1}^{lm}$ and $\sigma$ can provide us with the equivalent information as $1/T_{1}^{str}$ and $\beta$ obtained from the conventional stretched fit analysis.  

\begin{figure}
	\centering
	\includegraphics[width=3.2in]{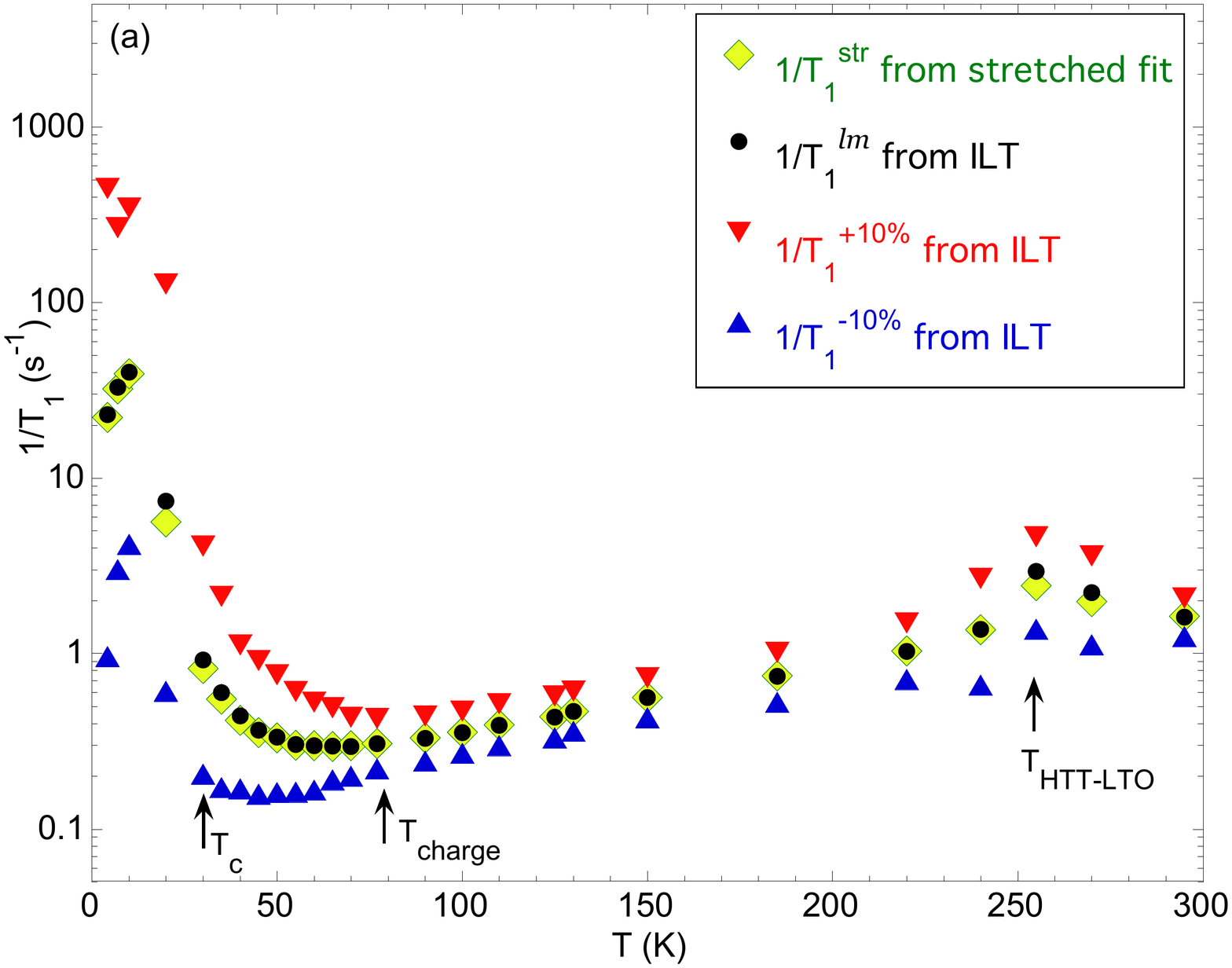}
	\includegraphics[width=3.2in]{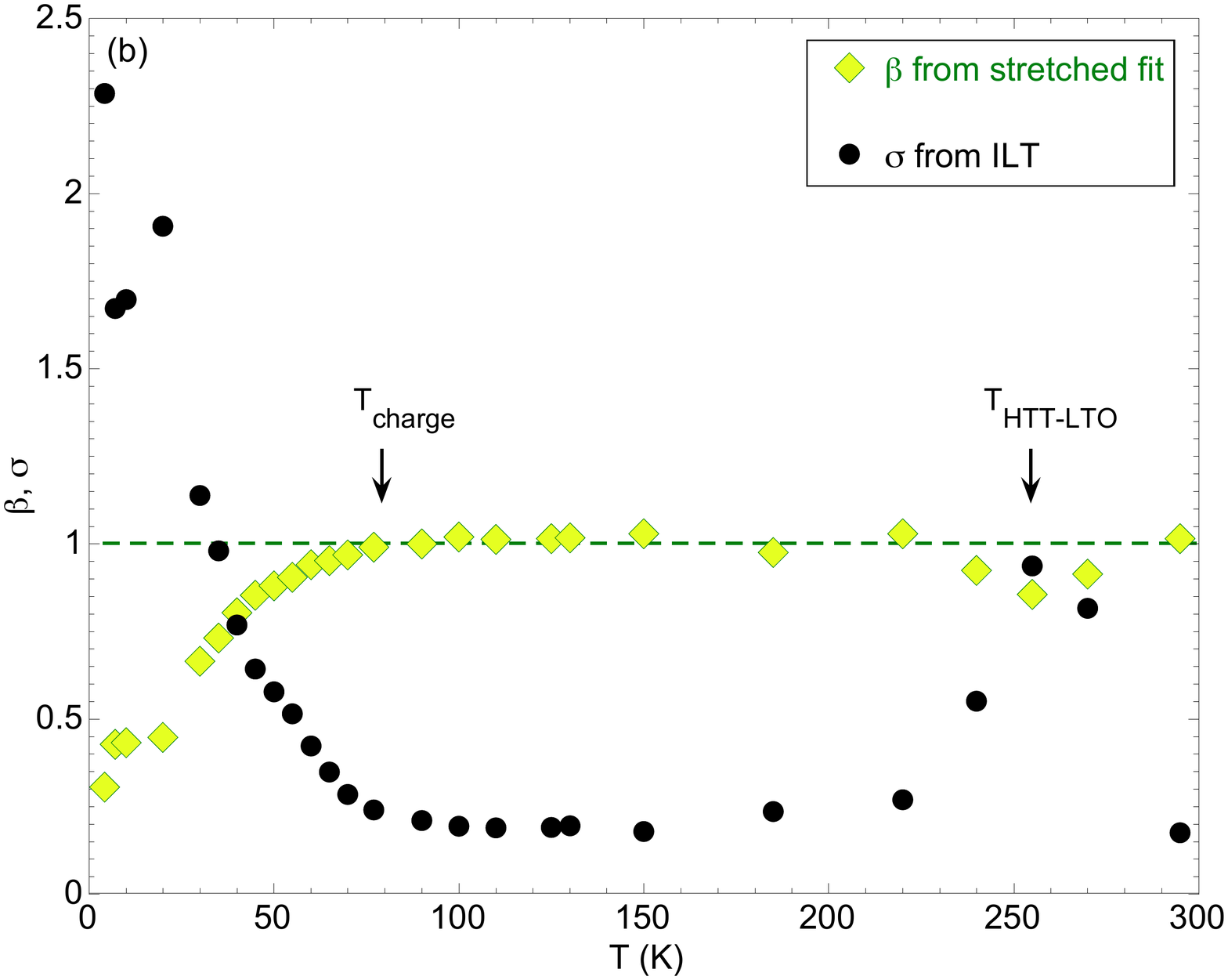}
	\caption{(a) Black filled bullets: $1/T_{1}^{lm}$, the log-mean value of the distributed $1/T_{1}$ estimated from $P(1/T_{1})$.  Red downward triangles: $1/T_{1}^{+10\%}$, the top 10\% value of the distributed $1/T_{1}$.  Blue upward triangles: $1/T_{1}^{-10\%}$, the bottom 10\% value of the distributed $1/T_{1}$.   Notice that $1/T_{1}^{+10\%}$ begins to grow as soon as charge order sets in, but $1/T_{1}^{-10\%}$ continues to slow down toward the simultaneous onset of superconductivity and magnetic order at $T_{c} \simeq T_{spin}^{neutron} \simeq 30$~K.   Also shown is $1/T_{1}^{str}$ (green diamonds), obtained from the single component stretched fit of $M(t)$ \cite{ArsenaultPRB2018}.  $1/T_{1}^{str}$ turns out to be a good approximation of $1/T_{1}^{lm}$, but fails to capture other important features in the distribution of $1/T_1$.  (b) The standard deviation $\sigma$ of $P(1/T_{1})$, in comparison to the stretched exponent $\beta$ for the stretched fit.  Notice the clear signature of anti-correlation.
	}
	\label{fig:hwhm}
\end{figure}

However, the ILT can go one more major step beyond elucidating the {\it average behavior}, and tell us {\it exactly how $1/T_1$ distributes about the average value $1/T_{1}^{lm}$}.  Let us now closely examine how charge order affects the distribution of the local magnetic environment.  First, note that $P(1/T_1)$ has a single narrow peak with minimal distributions between 295~K and $T_{charge}$ except near $T_{HTT-LTO}$.  This is consistent with the fact that the conventional stretched fit returns $\beta \simeq 1$ in this temperature range, implying that there is little distribution in $1/T_1$.  The finite width of $P(1/T_1)$ above $T_{charge}$ is set primarily by the resolution of the ILT \cite{SingerPRB2019}.  Also note that $1/T_{1}^{lm}$ is located at the peak of $P(1/T_1)$ down to $T_{charge}$, because the distribution of $P(1/T_1)$ is symmetrical except near $T_{HTT-LTO}$.  

The situation completely changes below $T_{charge} \sim 80$~K, as clearly shown by the close up view of  $P(1/T_{1})$ in Fig.\ 4.  $P(1/T_1)$ begins to broaden asymmetrically by extending a tail toward larger values of $1/T_1$ due to glassy spin freezing.  At 50~K and 45~K, the main peak of $P(1/T_1)$ is still shifting toward the smaller values of $1/T_1$, and the residue of the peak is recognizable as a shoulder at $1/T_{1} \sim 0.25$~s$^{-1}$ down to $\sim 20$~K.  These ILT results indicate that $1/T_1$ at a majority of $^{139}$La sites is still decreasing below $T_{charge}$, although the sample averaged $1/T_{1}^{lm}$ (filled bullets) is pulled toward the larger value of $1/T_1$ due to the growing tail toward larger $1/T_{1}$.  

To quantify these observations, in Fig.\ 3(a) we plot the top 10\% value $1/T_{1}^{+10\%}$ (red downward triangles) and the bottom 10\% value $1/T_{1}^{-10\%}$ (blue upward triangles) of distributed $1/T_1$, as estimated from $P(1/T_{1})$.  $1/T_{1}^{+10\%}$ and $1/T_{1}^{-10\%}$ keep track with the upper and lower bounds of the distribution in the color contour map in Fig.\ 2, as expected.  But $1/T_{1}^{+10\%}$ and $1/T_{1}^{-10\%}$ show qualitatively different behaviors below $T_{charge}$.  The faster parts in the sample, represented by $1/T_{1}^{+10\%}$, begin to show a divergent behavior as soon as charge order sets in.  On the other hand, the slower parts in the sample, represented by $1/T_{1}^{-10\%}$, continue to slow down through $T_{charge}$, as if nothing has happened, until divergent behavior finally sets in at $\sim 30$~K.  The temperature dependence of $1/T_{1}^{-10\%}$ observed above $T_c$ is very similar to the non-distributed $1/T_{1}$ observed above $T_c$ for the optimally superconducting phase La$_{1.85}$Sr$_{0.15}$CuO$_4$ \cite{Kobayashi, Yoshimura1992, BaekPRB2017}.  Combined with the continuing shift of the main peak of $P(1/T_{1})$ toward smaller values of $1/T_1$ in the shaded region of Fig.\ 4, we conclude that a significant volume of the CuO$_2$ planes continues to behave like the canonical high $T_c$ superconducting phase even below $T_{charge}$.  It is just that their volume fraction becomes smaller and smaller starting from $T_{charge}$, as evidenced by the suppression of the main peak of $P(1/T_{1})$.  $1/T_{1}^{str}$ appears to level off below $T_{charge}$, merely because it represents the average behavior of different domains with increasing $1/T_1$ and decreasing $1/T_1$.  In the case of La$_{1.875}$Ba$_{0.125}$CuO$_4$, $1/T_{1}^{str}$ \cite{HuntPRB2001, BaekLaT1PRB2015, SingerPRB2019} as well as $1/T_{1}^{lm}$,  $1/T_{1}^{+10\%}$ and $1/T_{1}^{-10\%}$ \cite{SingerPRB2019} begin to diverge at $T_{charge} \sim 54$~K, because charge ordered CuO$_2$ planes are more homogeneous.    

\begin{figure}
	\centering
	\includegraphics[width=3.2in]{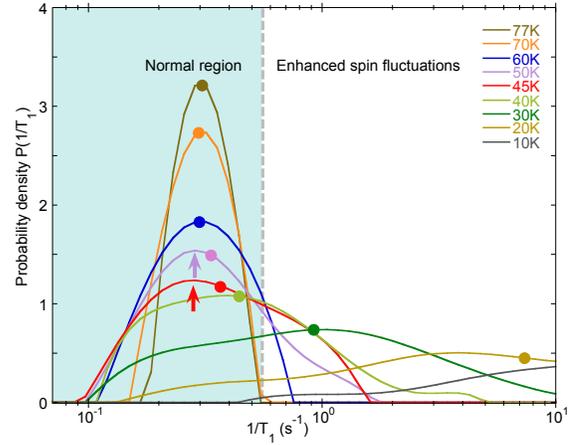}
	\caption{The closeup view of $P(1/T_{1})$ below $T_{charge} \sim 80$~K.  The filled bullets mark $P(1/T_{1}^{lm})$ at the average value $1/T_{1}^{lm}$.  Although $1/T_{1}^{lm}$ increases below 60~K, the most likely value of $1/T_{1}$ represented by the peak location (arrows) continues to slow down.  The corresponding shoulder is still recognizable at $1/T_{1} \sim 0.25$~$s^{-1}$ down to $\sim 20$~K.  The grey dashed vertical line represents the cut-off value $1/T_{1} = 0.55$~s$^{-1}$.  The integral of $P(1/T_{1})$ in the shaded region below the cut-off represents the volume fraction of the domains with slow $1/T_{1}$ expected for canonical superconducting CuO$_2$ planes, summarized with filled blue bullets in Fig.\ 5.   
	}
	\label{fig:recoveries-13}
\end{figure}

Our conclusion is consistent with the fact that the charge order correlation length is as short as $\xi \simeq3$~nm in  La$_{1.885}$Sr$_{0.115}$CuO$_4$ \cite{WenNatComm2019}.  In other words, the charge ordered phase is  highly disordered, and the properties of CuO$_2$ planes are expected to vary with the short length scale $\xi$.  Our conclusion is also corroborated by our earlier $^{63}$Cu NMR measurements conducted for the same piece of crystal \cite{ImaiPRB2017}.  A single, narrow $^{63}$Cu NMR peak observed above $T_{charge}$ is gradually wiped out, and the lost spectral weight is transferred to a much wider, wing-like signal that emerges below $T_{charge}$ underneath the narrow peak.  The wing-like signals have extremely fast NMR relaxation rates $1/T_1$ and $1/T_2$, and can be detected only with extremely short NMR pulse separation time $\tau \sim 2$~$\mu$s.  Since NMR is a local probe, the existence of these two different types of $^{63}$Cu NMR signals with drastically different relaxation rates indicates that there are two types of domains below $T_{charge}$ with different magnetic properties.  The $^{139}$La nuclear spins with slow $1/T_1$ in the blue shaded region of Fig.\ 4 and the narrow $^{63}$Cu NMR peak with slower relaxation rates originate from the same type of domains that are hardly affected by charge order.  The tailed section above the cut-off in Fig.\ 4 and the wing-like $^{63}$Cu NMR signals belong to the same domains, in which $1/T_1$ is locally enhanced due to spin freezing triggered by charge order.

\begin{figure}
	\centering
	\includegraphics[width=3.2in]{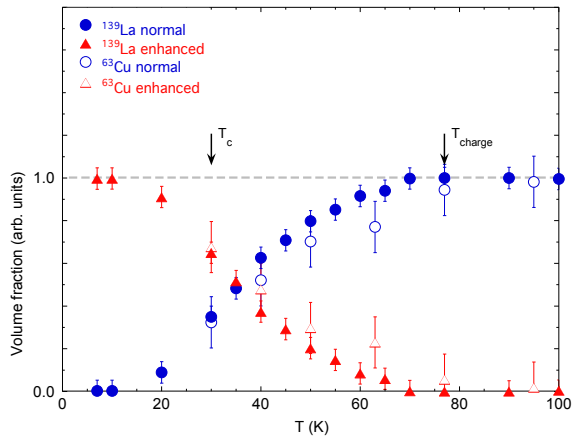}
	\caption{The volume fraction of the canonically superconducting CuO$_2$ planes with slow $1/T_{1}$, as estimated from the integral of $P(1/T_{1})$ in the shaded region of Fig.\ 4 below the cut-off (blue filled bullets), and from the intensity of the normal $^{63}$Cu NMR signal (blue open bullets \cite{ImaiPRB2017}).  Also plotted are the volume fraction of the charge ordered domains with enhanced $1/T_1$ estimated from the integral of $P(1/T_{1})$ above the cut-off (red filled triangles), and from the intensity of the wing-like $^{63}$Cu NMR signals (red open triangles \cite{ImaiPRB2017}).
	}
	\label{fig:t1}
\end{figure}

We can estimate the volume fraction of the domains affected by charge order using the probability density $P(1/T_{1})$, because its integral represents the number of $^{139}$La nuclear spins.  We introduce a cut-off at $1/T_{1} = 0.55$~s$^{-1}$ at the upper end of the distribution of $P(1/T_{1})$ observed at 77~K, see the grey dashed line in Figs.\ 1 and 4.  At 77~K, 100\% of the sample volume behaves similarly to canonically superconducting CuO$_2$ planes, and the entire $P(1/T_{1})$ curve is located below the cut-off.  Below $T_{charge}$, spin freezing sets in in charge ordered domains and $1/T_1$ increases at $^{139}$La sites in these domains.  As a consequence, a part of the $P(1/T_{1})$ curve extends above the cut-off.  The integrated area under the $P(1/T_{1})$ curve below the cut-off represents the fraction of the canonically behaving domains, whereas the integrated area above the cut off is from $^{139}$La nuclear spins in charge ordered domains.  Needless to say, the summation of the two volume fractions is 100\%, because the total integral of $P(1/T_{1})$ is normalized to 1. 

In Fig.\ 5, we summarize the temperature dependence of the volume fractions of charge ordered domains with enhanced Cu spin fluctuations (red filled triangles) and domains unaffected by charge order (blue filled bullets).  The agreement with our earlier estimation \cite{ImaiPRB2017} based on the relative $^{63}$Cu NMR intensities of the narrow peak and wing-like signals (open symbols) is satisfactory.  More importantly, notice that  at $T_{c}=30$~K a majority of $^{139}$La nuclear spins ($\sim 40$~\%) still relax with $1/T_1$ slower than the cut-off.  This means that glassy freezing of Cu spins triggered by charge order hardly affects $\sim 40$~\% of the CuO$_2$ planes when superconductivity sets in.  This finding is in remarkable contrast with the case of  La$_{1.875}$Ba$_{0.125}$CuO$_4$ with more strongly suppressed $T_{c}=4$~K \cite{SingerPRB2019, ImaiPRB2019}.  In La$_{1.875}$Ba$_{0.125}$CuO$_4$, the volume fraction affected by charge order quickly grows once charge order sets in at a much lower temperature $T_{charge} \sim 54$~K, and nearly 100~\% of the CuO$_2$ planes are covered by domains with highly enhanced spin fluctuations at 30~K.  These contrasting behaviors between La$_{1.885}$Sr$_{0.115}$CuO$_4$ and La$_{1.875}$Ba$_{0.125}$CuO$_4$ suggest that superconductivity with the higher onset temperature of $T_{c}=30$~K survives in the former, probably because nearly a half of the sample volume is unaffected by spin freezing at $T_{c}$ despite the higher onset of charge order.

\section{\label{sec:level1} Summary and Conclusions }
We have re-examined the $^{139}$La nuclear spin-lattice relaxation process in La$_{1.885}$Sr$_{0.115}$CuO$_4$ based on the inverse Laplace transform, and deduced the histogram of the domain-by-domain distribution of $1/T_1$ in the form of the probability density $P(1/T_{1})$.  We emphasize that we {\it deduced} $P(1/T_{1})$ by numerically inverting the experimentally observed recovery curve $M(t)$ without assuming the functional form of $M(t)$, such as the stretched exponential \cite{single} or double components \cite{double}.  

We demonstrated that the main peak of $P(1/T_{1})$, associated with the domains without enhanced spin fluctuations, persists even below $T_{charge}$.  At the onset of superconductivity at $T_{c}=30$~K, nearly a half of the sample volume remains unaffected by charge order, in agreement with our earlier $^{63}$Cu NMR results.  

Our work reported here and elsewhere \cite{SingerPRB2019, TakahashiPRX2019} illustrates the powerful nature of the new ILT techniques applied to disordered quantum materials.  The ILT not only probes the average behavior of distributed $1/T_1$ in the form of $1/T_{1}^{lm}$, but also deduces the histogram of distributed $1/T_1$.  The conventional stretched fit analysis of $M(t)$ discards the information about the latter.  

\section{Acknowledgement}
We thank S.\ K.\ Takahashi and J.\ Wang for their assistance in NMR data acquisition and analysis.  T.I is financially supported by NSERC.  P.M.S. is supported by The Rice University Consortium for Processes in Porous Media.  The work at Tohoku was supported by Grant-in-Aid for Scientific Research (A) (16H02125), Japan.



%

\end{document}